# A Multi-Moded RF Delay Line Distribution System for the Next Linear Collider[*]

S. G. Tantawi[†], C. Nantista, N. Kroll, Z. Li, R. Miller, R. Ruth, P. Wilson, Stanford Linear Accelerator Center, SLAC, 2575 Sand Hill Rd. Menlo Park, CA 94025, USA

and J. Neilson, Calabazas Creek Research, Inc., 20937 Comer Drive, Saratoga, California 95070

**Abstract.** The Delay Line Distribution System (DLDS) is an alternative to conventional pulse compression, which enhances the peak power of rf sources while matching the long pulse of those sources to the shorter filling time of accelerator structures. We present an implementation of this scheme that combines pairs of parallel delay lines of the system into single lines. The power of several sources is combined into a single waveguide delay line using a multi-mode launcher. The output mode of the launcher is determined by the phase coding of the input signals. The combined power is extracted from the delay line using mode-selective extractors, each of which extracts a single mode. Hence, the phase coding of the sources controls the output port of the combined power. The power is then fed to the local accelerator structures. We present a detailed design of such a system, including several implementation methods for the launchers, extractors, and ancillary high power rf components. The system is designed so that it can handle the 600 MW peak power required by the NLC design while maintaining high efficiency.

## I. INTRODUCTION

During the past few years, high-power rf pulse compression systems have developed considerably. These systems provide a method for enhancing the peak power obtainable from the output of high power rf sources. One important application is driving accelerator structures. In particular, future linear colliders, such as the proposed NLC, require peak rf powers that cannot be generated by the current state-of-the-art microwave tubes. The SLED pulse compression system [1] was

---

[*] Work supported by Department of Energy contract DE-AC03-76SF00515.
[†] *Also with the Electronics and Communications Department Cairo University, Giza, Egypt.*

Submitted to Physical Review Special Topics – Accelerators and Beams, January 2002.

implemented to increase the accelerating gradient of the two-mile linac at the Stanford Linear Accelerator Center (SLAC).

One drawback of SLED is that it produces an exponentially decaying output pulse. Approaches to shaping the pulse by means of phase or amplitude modulation of the input pulse [2,3,4] could obtain only modest efficiency. To produce a flat pulse and improve efficiency, the Binary Pulse Compression (BPC) system [5] was invented. The BPC system has the advantage of 100% intrinsic efficiency and a flat output pulse. Also, if one accepts some efficiency degradation, it can be driven by a single power source. However, The implementation of the BPC [6] requires a large assembly of over-moded waveguides, making it expensive and extremely large in size. The SLED II pulse compression system is a variation of SLED that gives a flat output pulse [7]. The SLED II intrinsic efficiency is better than that of SLED, but not as good as that of BPC. However, from the compactness point of view SLED II is far superior to BPC. Several attempts have been made to improve its efficiency by turning it into an active system [8]. However, the intrinsic efficiency of the active SLED-II system is still lower than that of the BPC.

The DLDS [9] is a system similar to BPC that utilizes the delay of the electron beam in the accelerator beam line of the linear collider to reduce the length of the over-moded waveguide assembly. However, it still uses more over-moded waveguide than that required by SLED II. To further enhance the DLDS we introduce in this paper a variation on that system which further reduces the length of the waveguide system by multiplexing low-loss rf modes in the same waveguide, hence the name Multi-moded

DLDS (MDLDS). This system has an intrinsic efficiency of 100%, and, assuming a power-handling limit of 600 MW, requires no more delay line per feed than dual-moded SLED-II (half that of single-moded SLED-II).

In section II, we present a basic description of the system, its function, and our approach to component design. In section III, we discuss the timing issue of keeping the rf synchronized with the propagating electron (or positron) beam. In section IV, we describe the design and implementation of the multi-mode launcher, including mode converting cross-section tapers. In section V, we describe some ancillary components, such as bends. In section VI, we illustrate the integration of components into a combining/launching circuit. In section VII, we describe the mode-selective extractor and tap-offs for distributing power from a given waveguide to the appropriate structures. We conclude in section VIII.

## II. SYSTEM DESCRIPTION

In a DLDS system, groups of klystrons are made to deliver their combined power to a sequence of distantly separated accelerator feeds during consecutive time-bin divisions of the full operating pulse. Power direction is accomplished by switching the relative phases of the rf drives. With a proper passive, matched circuit, four sources, for example, can be made to feed four feeds through the four orthogonal phase combinations. During the first time-bin, power is shipped furthest upstream; during the last, it is delivered locally. The shortened propagation distance of the rf combines with the later arrival of the beam to allow time for the structures powered by each consecutive feed to be filled. With MDLDS, we eliminate the need for a separate delay line to transport the rf to each feed. Power is delivered to two or more feeds through the same waveguide by using different modes as the carriers and inserting specially designed extractors to direct each mode to its proper destination.

The modes of choice in our circular waveguide delay lines are the low-loss $TE_{01}^\circ$ mode and the $TE_{12}^\circ$ mode[*]. The $TE_{12}^\circ$ mode becomes highly efficient as we increase the waveguide diameter. At a diameter of approximately five times the free space wavelength, this mode has the lowest attenuation after the well-known $TE_{01}^\circ$ mode. Recently this has been verified experimentally [10].

The initial three-mode plan for the MDLDS system utilized the $TE_{01}^\circ$ mode and both polarizations of the $TE_{12}^\circ$ mode. Component designs for that system involved coupling slots and irises, which had the potential to occasion rf breakdown limitations. As our system must be able to carry up to 600 MW, we have more recently concentrated our efforts on a more conservative two-mode approach in which we manipulate the rf in moderately overmoded, rectangular waveguide. Transitioning to and from the highly overmoded circular delay line waveguide will be done through special mode-order-preserving tapers. Current component designs exploit planar symmetry, which allows for more facile mode manipulation and for the use of waveguide of arbitrary height to limit field levels. Our goal has been to keep the electric field below ~40 MV/m while at the same time aiming for compactness to minimize ohmic losses.

---

[*] We use a superscripted circle with circular waveguide modes to avoid confusion with rectangular waveguide modes, to which this paper will also refer.

Figure 1 gives a schematic of a module of our MDLDS, whose components will be described below. Eight 75 MW, 11.424 GHz klystrons are combined in pairs, yielding four independently phased sources. The combined 600 MW, 1.5 µs pulse is phased into four 375 ns time bins. The feeds are numbered to show the order in which they receive power. Each feed delivers a nominal 200 MW to each of three structures. A number of interleaved modules power the intervening structures.

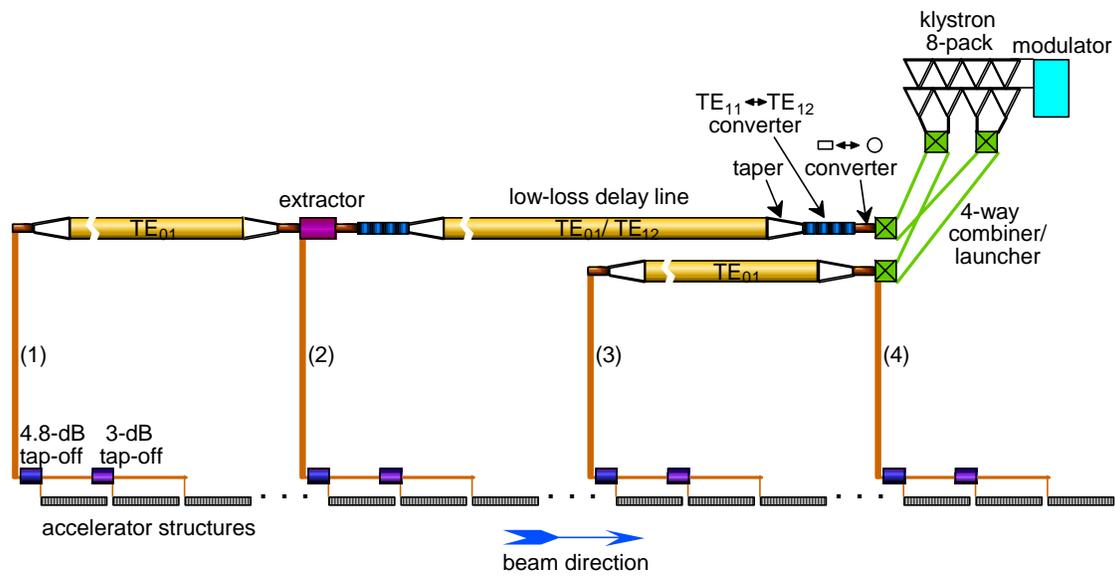

**FIGURE 1.** Schematic of the dual-moded MDLDS with four accelerator feeds.

## III. TIMING

Because the rf power is being injected at different times into different modes that have different group velocities, one must pay special attention to timing. The set of equations that needs to be satisfied so that each group of accelerator structures gets an rf pulse for a duration $\tau$ at the appropriate time is:

$$\tau = (\frac{L_1}{v_{TE01}} + \frac{L}{c}) + (\delta_3 - \delta_4),$$

$$\tau = (\frac{L_2}{v_{TE12}} - \frac{L_1}{v_{TE01}} + \frac{L}{c}) + (\delta_2 - \delta_3), \qquad (1)$$

$$\tau = (\frac{L_3}{v_{TE01}} + \frac{L}{c}) + (\delta_1 - \delta_2) + L_2(\frac{1}{v_{TE01}} - \frac{1}{v_{TE12}}),$$

where $L$ is the beam line spacing between accelerator structure groups, $L_1$ is the distance between the launcher and group (3), $L_2$ is the distance between the launcher and the extractor, $L_3$ is the length of the delay line after the extractor, $v_{TE01}$ and $v_{TE12}$ are the group velocities of the $TE_{01}°$ and $TE_{12}°$ modes respectively, and $\delta_1$ through $\delta_4$ are the delays due to the transmission of power from the main rf delay line system to the accelerator structure groups, i.e., the delay through the extractors and feeds.

There are several choices for the lengths $L$, $L_1$ through $L_3$, and $\delta_1$ through $\delta_4$ that satisfy the above set of equations. An attractive choice is to set $L_1$ through $L_3$ equal to $L$, and

$$\delta_1 = \delta_3 = \delta_4 + \frac{L}{2}(\frac{1}{v_{TE12}} - \frac{1}{v_{TE01}})$$

$$\delta_2 = \delta_4 - L(\frac{1}{v_{TE12}} - \frac{1}{v_{TE01}}). \qquad (2)$$

Then, the distance between structure groups fed by a given module would be given by

$$L = \frac{\tau}{(\frac{1}{2v_{TE01}} + \frac{1}{2v_{TE12}} + \frac{1}{c})}. \qquad (3)$$

This would lead to a fairly symmetric system.

# IV. LAUNCHER

*A. Representation*

Several ideas for the launcher have been proposed [11,12]. In all of them, a fundamental property of the launcher has been preserved: the launcher has only four inputs, and the launcher has to launch four and only four modes. If such a device is matched for all four orthogonal input conditions, because of unitarity and reciprocity, the scattering matrix representing the launcher has to take the following form:

$$S = \begin{bmatrix} 0 & 0 & 0 & 0 & 1/2 & 1/2 & 1/2 & 1/2 \\ 0 & 0 & 0 & 0 & -1/2 & -1/2 & 1/2 & 1/2 \\ 0 & 0 & 0 & 0 & -1/2 & 1/2 & -1/2 & -1/2 \\ 0 & 0 & 0 & 0 & -1/2 & 1/2 & -1/2 & 1/2 \\ 1/2 & -1/2 & -1/2 & -1/2 & 0 & 0 & 0 & 0 \\ 1/2 & -1/2 & 1/2 & 1/2 & 0 & 0 & 0 & 0 \\ 1/2 & 1/2 & 1/2 & -1/2 & 0 & 0 & 0 & 0 \\ 1/2 & 1/2 & -1/2 & 1/2 & 0 & 0 & 0 & 0 \end{bmatrix}. \quad (4)$$

This form forces isolation between inputs; i.e., none of the input ports will receive any reflected power as a result of one or more sources dropping out or failing. If one of the four power sources fails (two klystrons are off), the available power at the intended destination will be reduced by a factor of 9/16, and the total acceleration by 3/4. If one klystron in any pair fails, half of the remaining klystron's power will go to a load on the fourth port of the hybrid through which they are combined, so the available input to the launcher will be reduced by a factor of 1/4. The total combined power is then reduced by a factor of 49/64, and the acceleration by a factor of 7/8. More generally, the power available at the structures is proportional to the square of the fraction of operating klystrons, and the gradient is proportional to that fraction.

The fraction of the power that is misdirected as a result of klystron failure is distributed between loads and the wrong structure sets for each time bin.

*B. Cross Potent Superhybrid/Launcher*

The move in our pulse compression work toward rectangular waveguide components, in which planar symmetry is exploited to allow arbitrary height, began with the design of planar hybrids [13] to replace magic T's in the SLED-II system of the ASTA (Accelerator Structure Test Area) facility at SLAC. These had been exhibiting rf breakdown problems above 200 MW, particularly at the mouth of the E-plane port. One novel hybrid design has an "H" geometry (see Figure 11). Its central guide is wide enough to support two TE modes, and, at its junctions, triangular wall protrusions yield, essentially, double mitred bends. Prototypes have been built and successfully operated at peak power levels approaching 500 MW.

If two such "magic H" hybrids, with ports half the width of the central guide, are placed side-by-side and their common wall removed, the resulting oversized ports have the same cross-section as the central guide. If these are split again with T's at the proper distance, the symmetry is completed, and an eight port device in the shape of a cross potent (a heraldic cross with a bar at each extremity) results. This "cross potent superhybrid" can be used to combine power from four input ports, by proper phasing, into any one of four output ports. Opposite pairs of cross arms are isolated. A prototype has been built and its scattering parameters measured with a network analyzer, with very satisfactory results [14].

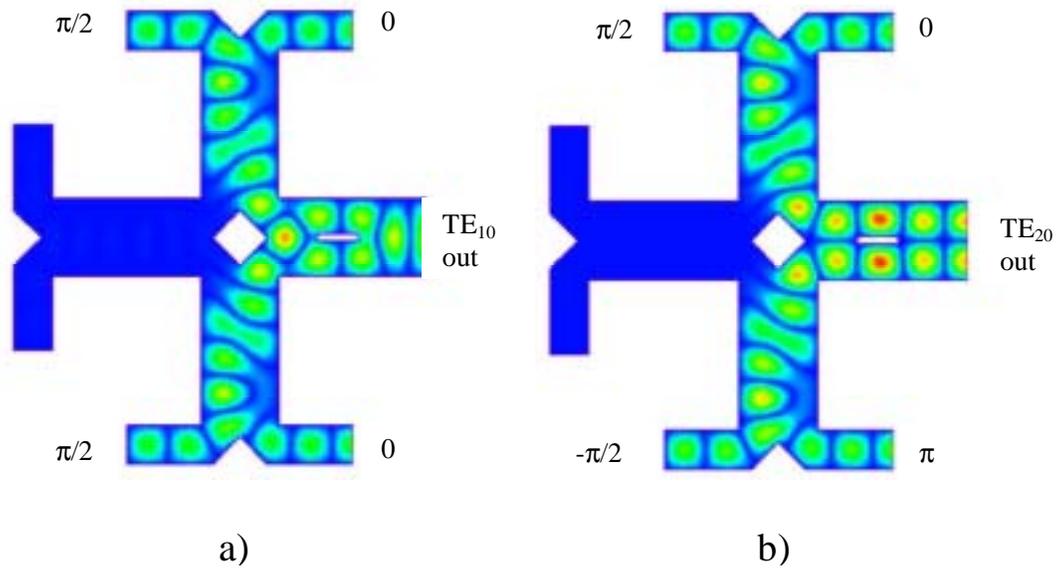

**FIGURE 2.** Cross Potent Launcher with simulated electric field plots illustrating launching a) $TE_{10}$ and b) $TE_{20}$ in the right overmoded rectangular port with the indicated relative phases for four equal amplitude inputs. Alternate phasings of the inputs send the power to either of the left ports.

Of course, one can leave off the T split on one or more of the arms, substituting posts for matching, and consider the rectangular $TE_{10}$ and $TE_{20}$ modes as the orthogonal outputs or inputs for such arms. This cross potent launcher configuration and its function are illustrated in the HP HFSS [15] simulation field plots of Figure 2. The wide waveguide width is 1.442 inches, and the height can be chosen to accommodate the taper/converters described below. Combined with these taper/converters, it will allow the desired modes to be launched into circular waveguide delay lines from four independently phased sources.

*C. Circular-to-Rectangular Taper Converter*

One can generate $TE_{12}^o$ from $TE_{11}^o$ via an adiabatically corrugated, or rippled, circular waveguide. The wall undulations can be designed to pass $TE_{01}^o$ unperturbed. The challenge then becomes to launch $TE_{01}^o$ and $TE_{11}^o$ into the same waveguide and to extract one while passing the other.

The components to be described throughout this paper perform these functions with the rectangular $TE_{10}$ and $TE_{20}$ modes. We can use them in our system without sacrificing the benefits of circular waveguide delay lines if we can transition between the two cross-sections in such a way that a one-to-one correspondence is achieved for the respective operating modes. An adiabatic cross-section taper naturally converts $TE_{10}$ to $TE_{11}^o$. $TE_{20}$, however, tends to produce a combination of the circular modes $TE_{21}^o$ and $TE_{01}^o$. The cross-section deformation must be done in two or more properly designed and spaced taper sections to yield finally a pure $TE_{01}^o$ wave [16]. A taper-mode transducer accomplishes this in the space of a few inches without compromising the $TE_{11}^o$ conversion. This design is illustrated in Figure 3 with electric field plots from an HP HFSS simulation. The parasitic mode loss is negligible. (Since the rectangular to circular transition preserves mode order and the $TE_{01}^o$ cutoff is above that of $TE_{21}^o$, it is technically the higher cutoff polarization, $TE_{02}$, which produces $TE_{01}^o$. However, a slight width taper, necessary to match into the following components, makes the dimension along which the field varies larger, correcting this nomenclature at the port.) The ending diameter is 1.6 inches. The $TE_{11}^o$- $TE_{12}^o$ conversion and transition to low-loss 4.75 inch diameter guide is made through a taper design described below which also preserves $TE_{01}^o$.

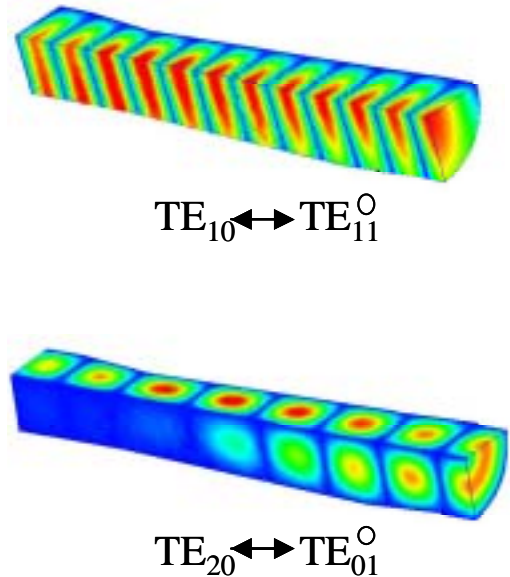

$TE_{10} \leftrightarrow TE_{11}^O$

$TE_{20} \leftrightarrow TE_{01}^O$

**FIGURE 3.** Taper Converter ¼ geometry, showing transition from rectangular to circular cross-section, with field patterns illustrating conversion between $TE_{10}$ and $TE_{11}^o$ and between $TE_{20}$ and $TE_{01}^o$.

*D. Diameter Taper/Mode Converter*

The design of this component is more challenging than a typical converter taper design in that it must convert the $TE_{11}^o$ to the $TE_{12}^o$ mode with at least 99% conversion efficiency while passing the $TE_{01}^o$ mode with minimal mode conversion. The component is also operated bi-directional, that is, it must convert a $TE_{12}^o$ mode at the output back into a $TE_{11}^o$ mode. In addition, the converter-taper combination should have reflections below -26 dB for both the $TE_{11/12}^o$ mode and the $TE_{01}^o$ mode.

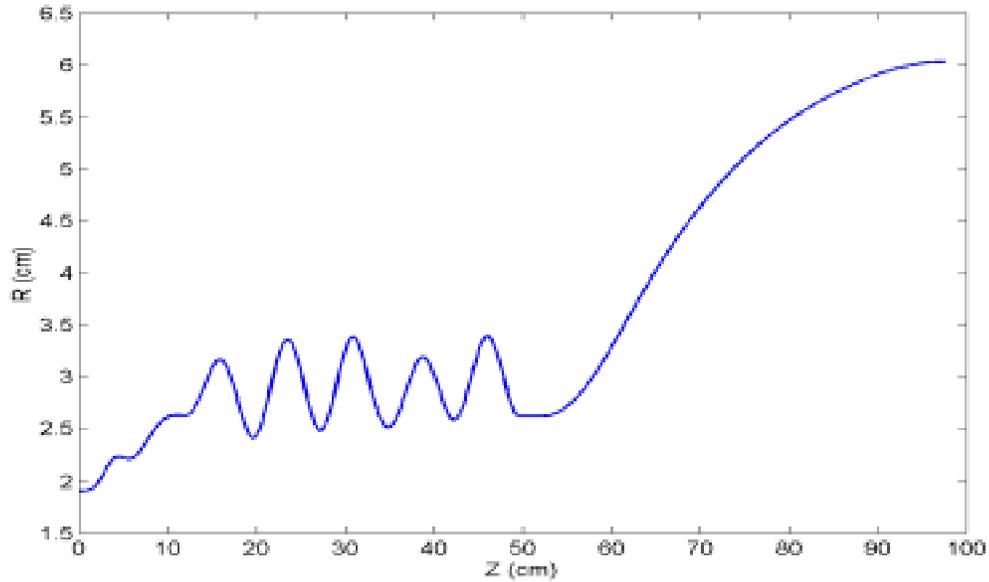

**FIGURE 4.** Profile of Diameter Taper-Converter.

The converter and taper profiles were all optimized for mode purity using a cubic spline fit of selected points along an initial guess for the wall profile. Using a cubic spline fit [17] (with zero derivative end conditions) ensured that the resulting wall profile varied smoothly so as to minimize unwanted reflections. The initial trial shape for optimization was a sinusoid with a period equal to the beat wavelength of the $TE_{11}°$ and $TE_{12}°$ modes. The total length of the system consisting of an initial up-taper followed by the converter and a final taper was 97.5 cm. The final converter taper had a mode purity of 99.8% for $TE_{12}°$ and 99.9% for $TE_{01}°$, and the all return loss factors were less than -30 dB. A plot of the taper profile is shown in figure 4.

## V. BENDS AND OTHER COMPONENTS

The physical layout of our rf system will require bends which, because the system is overmoded, are not completely trivial. Even at places where a single mode is used, power levels generally do not allow us to reduce the cross-section back to single-moded waveguide. We now describe plans for negotiating such bends and otherwise manipulating the rectangular modes.

*A. Overmoded H-plane Bend*

We have designed a 90º H-plane bend in the 1.442 inch-wide waveguide mentioned above. Just as we found a radius-of-curvature that gave 50% conversion at 45º for the extractor, one expects that certain bending radii will bring power coupled along the bend completely back to the entering mode. We find that, for this cross-section at our operating frequency, a bend with a radius-of-curvature from the inner wall of 1.409 inches transmits each of the two operating modes with essentially 100% purity from input to output, as shown in Figure 5. As with the above planar components, other propagating modes, including TM modes, are not coupled in these bends. Since reflections are negligible and the coupling is only between $TE_{10}$ and $TE_{20}$, the solution automatically works for both modes.

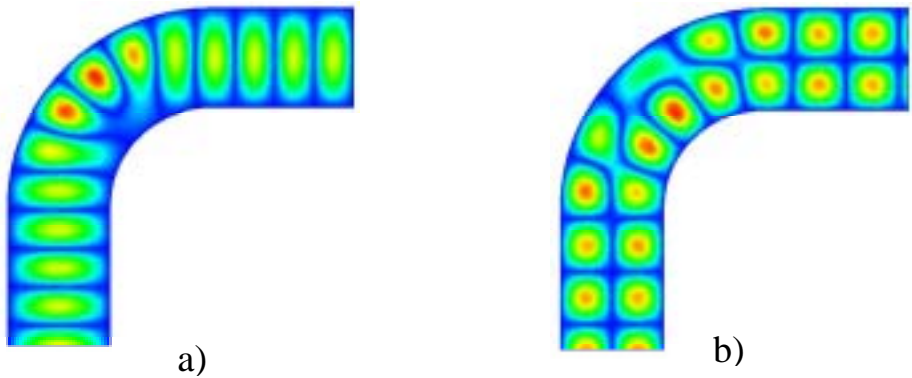

a)  b)

**FIGURE 5.** Overmoded H-plane bend in 1.442" waveguide with simulated electric field plots illustrating a) $TE_{10}$ mode transmission and b) $TE_{20}$ mode transmission.

*B. Jog Converter*

A 45º H-plane bend with an inner-wall radius-of-curvature of 1.055 inches first mixes the two rectangular guide modes, converting either input into an equal combination of $TE_{10}$ and $TE_{20}$. A second such bend may be located so as to either return the coupled power to the original mode or transfer the remaining power from it. This second bend can be in the same or the opposite sense, resulting in either a 90º bend or a jog.

The "jog converter" is a compact mode transducer, consisting of two oppositely oriented 45º bends, separated by a very short phasing section, which gives complete conversion between $TE_{10}$ and $TE_{20}$. It works in either direction, for either input mode. This simple device is shown in Figure 6. It is used at several points in our rf system plans. It can be combined with a rectangular waveguide taper and a rectangular-to-circular taper converter of the type described above to form a novel $TE_{10}$ to $TE_{01}º$ launcher.

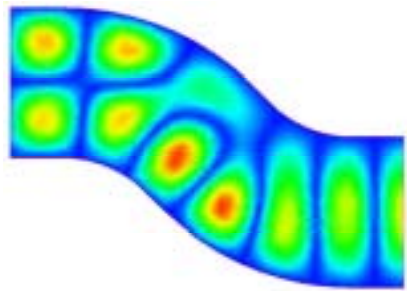

**FIGURE 6.** Jog converter geometry with simulated electric field plots illustrating conversion from $TE_{20}$ to $TE_{10}$ (left to right) or from $TE_{10}$ to $TE_{20}$ (right to left).

*C. Overmoded E-plane Bend and Height Taper*

There are points in our rf system where components are required for which planar symmetry cannot be maintained. Two rectangular waveguide components that fall into this category are E-plane bends and height tapers.

In our power-combining waveguide configuration, a few E-plane bends are unavoidable. It is desirable, for power handling and matching into other components, that these be in full overmoded height. Fortunately, they needn't be in our full overmoded width; thus, only $TE_{10}$ must be transmitted. This mode is coupled strongly to $TM_{11}$ in such a bend. At a 0.900"×1.435" cross-section, an inner (bottom) wall radius-of-curvature of 1.510 inches returns a quite pure $TE_{10}$ mode (99.96%) at 90º, as seen in Figure 7a). The surface field is enhanced within the bend, but the maximum power it needs to handle in our system is only 75 MW.

While the above bend and parts of other components propagate a single $TE_{n0}$ mode, it has been assumed that the rectangular guide components described above will be built with an overmoded height (allow propagation of modes with non-zero second index). At points in the rf system, specifically at the structure inputs, it will be necessary to transition to true single-moded waveguide. Figure 7b) shows a preliminary double-stepped height taper design shown in going from 0.900"×0.400" (WR90) to 0.900"×1.210" in less than an inch with perfect transmission. However, even with well-rounded edges, the field enhancement gives ~70 MV/m at 200 MW. Thus, while an adiabatic three inch linear taper can be added to bring this device to the final 1.435 inch height while minimizing mode conversion to $TM_{12}$ and $TE_{12}$, we plan to develop a more compact design with lower surface field.

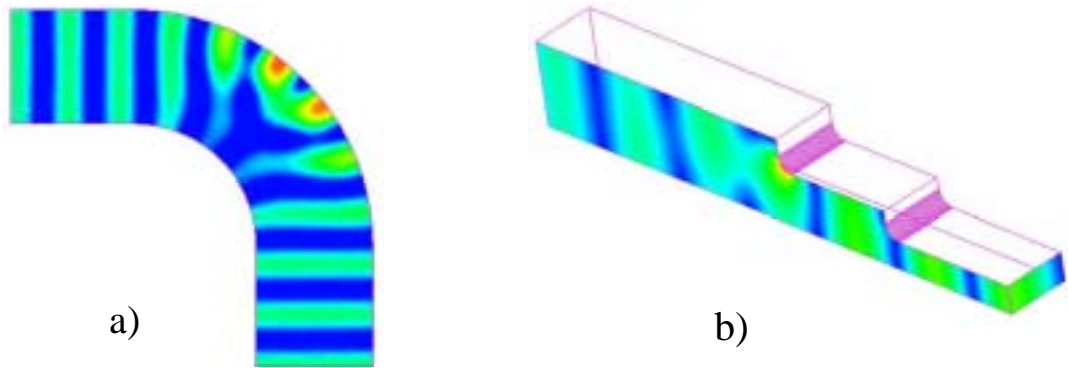

**FIGURE 7.** Simulated electric field plots in a) overheight E-plane bend and b) quarter geometry of a height taper.

## VI. THE 4-WAY COMBINER/LAUNCHER CIRCUIT

The launcher circuit can be constructed using the waveguide components described above. Figure 8 illustrates a way of combining power through eight input ports into either mode in a dual-moded delay line (1), one mode in a second delay line (2), or a local feed (3).

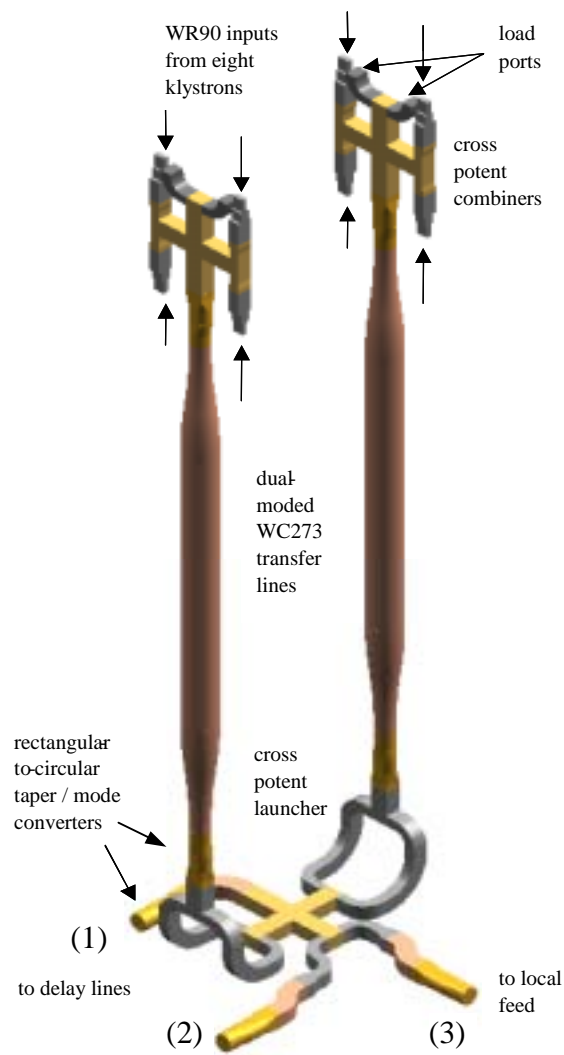

**FIGURE 8.** 8x4 combiner/launcher waveguide circuit for DLDS head.

The cross-potent superhybrid is used three times. At the first level, signals from each set of four klystrons are combined, using superhybrids, and directed to one of two modes in a transfer waveguide. On the second-level, the mode used in both transfer lines determines which way the power turns at the center of the superhybrid, and the relative phase of the two signals determines where it goes from there.

Note that using this configuration, rather than combining pairs of klystrons for four separate feeds to the final cross potent, allows the power to be fed from the first level to the second level in two dual-moded waveguides. This halves the number of transfer lines, required under the assumption that the klystron gallery and delay line installation are physically remote (the latter presumably being in the accelerator tunnel).

For the E-plane bends shown, the power is first split by T-junctions, matched for both modes, and later recombined in the same way. This allows use of the half-power, $TE_{10}$ bends described above. It also provides an opportunity to introduce a $\pi/2$ phase length difference in the two arms, required for power direction in the cross potent superhybrid.

## VII. EXTRACTOR AND TAP-OFFS

*A. Mode-Selective Extractor*

One system feed spacing before the end of a dual-moded delay line, one mode is extracted and fed into the accelerator. Following the proper tapers and converters, the extractor begins in the same rectangular dimensions as the launcher output. A 45° bend like that at the beginning of the jog converter leaves power entering in either operating mode in an equal mixture of the two. A short straight section is used to achieve the proper relative phase. Then a doubly matched T split, at which the $TE_{10}$ field adds constructively to one lobe of $TE_{20}$ and destructively to the other, sends all the power one way for a given extractor input mode and all the power the other way for the other input mode. Since the two input modes result in combinations with opposite relative phases, they excite opposite ports at the split. Again, we illustrate the geometry and function with field plots in Figure 9. Since it corresponds to the circular delay line mode with the greater attenuation, the $TE_{10}$ mode is selected for extraction.

Single-moded 45º H-plane bends orient the extraction port waveguide perpendicular to the delay line and the through port waveguide parallel to it, although slightly offset. The latter is then tapered to full width and sent through a dogleg or jog converter, described above, which simultaneously brings the port back in line with the delay line axis and restores the $TE_{20}$ mode. An identical mode converter can be appended to the extraction port, so that, through rectangular-to-circular taper converters, the power will be re-launched in either the delay line or the accelerator feed in the more efficient circular $TE_{01}º$ mode. By reciprocity, power reflected from either output port will be returned to the input port in the mode in which it arrived.

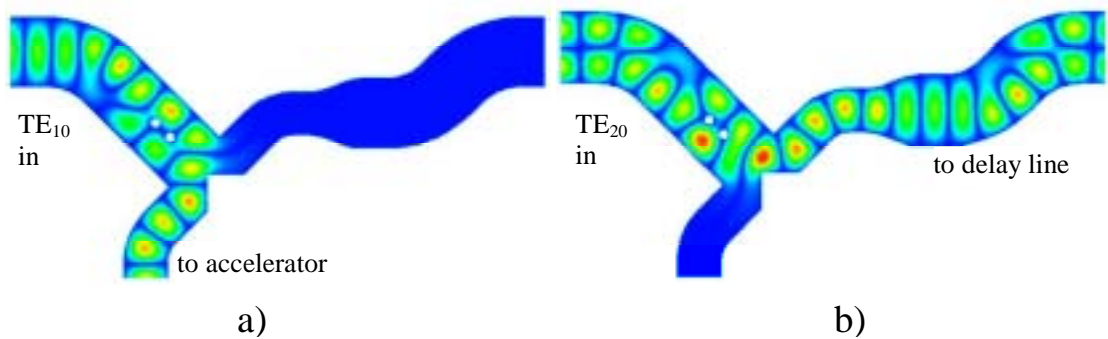

**FIGURE 9.** Extractor with simulated electric field plots illustrating a) extraction of the $TE_{10}$ mode and b) passing the $TE_{20}$ mode. For the latter, a "jog converter" attached to the through port after a width taper restores the mode and brings the port back in line with the delay line.

## B. PowerDistributing Tap-Offs

Each feed from the MDLDS shown in Figure 1 delivers ~600 MW to a set of three consecutive accelerator structures. After the waveguide turns to run parallel to the accelerator, first one third of the power will need to be removed for the first structure, and then one half of the remaining power will need to be removed for the second structure before the feed terminates in the third. We plan to do this as well in overmoded planar rectangular waveguide components.

One idea is to simply peel off a lobe from the appropriate $TE_{n0}$ mode for each tap-off. The $TE_{30}$ mode can be generated in widened waveguide from $TE_{10}$ or $TE_{20}$ by a planar converter (see for example [18], pg. 6). One third of the waveguide could then be interrupted by a mitred bend, which leaves two thirds of the field pattern and of the power to form a $TE_{20}$ wave in the continuing guide. At the second structure, one half of a $TE_{20}$ wave is similarly diverted. This concept for a 4.77-dB and a 3.01-dB power divider is illustrated in Figure 10.

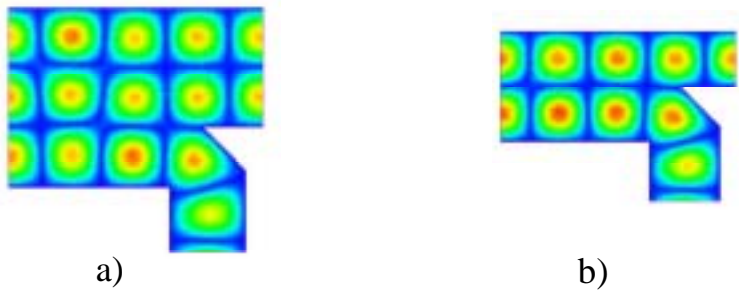

a)                                           b)

**FIGURE 10.** Tap-offs utilizing the idea of achieving fractional power division by deflecting a lobe of a $TE_{n0}$ field pattern. This requires converting to $TE_{30}$ before the first of a structure triplet.

To isolate the structures in case of rf breakdown, it may be preferable to use directional couplers instead. A hybrid of the "magic H" type on which the cross potent launcher is based can serve for the second tap-off. This is illustrated in Figure 11. For the first tap-off, the hybrid design can be modified to give the proper 1/3-2/3 split. This is simply a matter of adjusting the differential phase length of the coupling section while maintaining the match. Any reflected power from the structures would then travel back through the waveguide system or into high power loads on the fourth ports of the directional couplers. A second design with reduced coupling can serve for the 1/3 power tap-off. Jog converters may be used here, as shown, if the accelerator structure length warrants going to $TE_{01}$ in between structures for reduced losses.

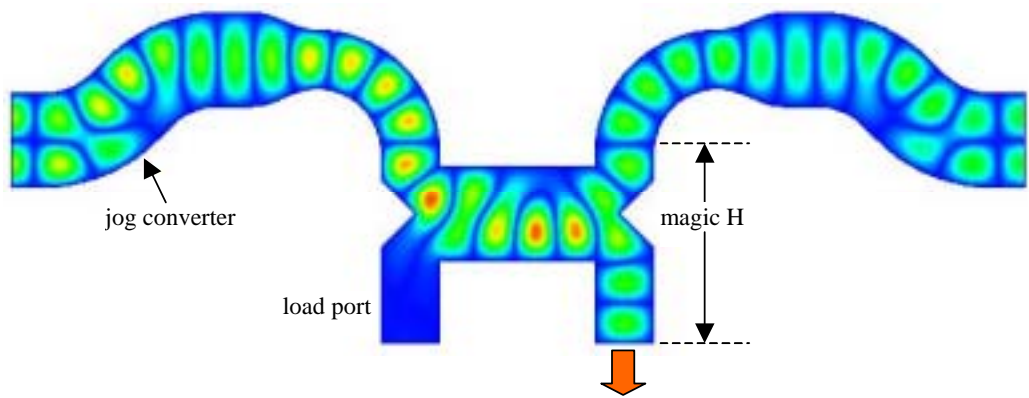

**FIGURE 11.** A 3-dB directional coupler tap-off with simulated field plots. Power flow is from left to right.

As the structure inputs will be spaced about two meters apart, it is worthwhile to convert back to the circular $TE_{01}°$ mode in between them. Taper converters combined with jog converters will get us to and from the proper rectangular mode.

## VIII. CONCLUSIONS

We have described a dual-moded MDLDS rf waveguide system for the NLC and several novel components designed to accomplish the various required functions. The relatively open geometry and exploitation of planar symmetry in our design motif allows us to keep the peak surface fields at reasonable levels for 600 MW at X-band. The use of highly-overmoded, circular waveguide is maintained for low-attenuation in delay lines and power transport, while rectangular waveguide is used in moderately overmoded components for the flexibility and simplicity of design provided by reduction to two dimensions. The current two-mode scheme requires one third less delay line than a simple four-feed DLDS. The designs described here might be suitably scaled for operation at 30 GHz or higher to serve in a pulse compression system for a higher frequency linear collider or for any other rf system with high power and small bandwidth requirements. Their simplicity may recommend them for use at low power as well.

## ACKNOWLEDGMENT

This work is supported by Department of Energy Contract DE-AC03-76SF00515.